# Chip-compatible quantum plasmonic launcher

Chin-Cheng Chiang*, Simeon I. Bogdanov, Oksana A. Makarova, Xiaohui Xu, Soham Saha,

Deesha Shah, Di Wang, Alexei S. Lagutchev, Alexander V. Kildishev,

Alexandra Boltasseva, and Vladimir M. Shalaev**

School of Electrical & Computer Engineering and Birck Nanotechnology Center, Purdue University, West Lafayette, IN 47907, USA

Purdue Quantum Science and Engineering Institute, Purdue University, West Lafayette, IN 47907, USA

*E-mail: chiang53@purdue.edu

**E-mail: shalaev@purdue.edu

Keywords: plasmonic launcher, in-plane emission, deterministic assembly, quantum plasmonics, nitrogen-vacancy centers, single-photon source, nanodiamonds

**Abstract**

Integrated on-demand single-photon sources are critical for the implementation of photonic quantum information processing systems. To enable practical quantum photonic devices, the emission rates of solid-state quantum emitters need to be substantially enhanced and the emitted signal must be directly coupled to an on-chip circuitry. The photon emission rate speed-up is best achieved via coupling to plasmonic antennas, while on-chip integration can be easily obtained by directly coupling emitters to photonic waveguides. The realization of practical devices requires that both the emission speed-up and efficient out-coupling are achieved in a single architecture. Here, we propose a novel platform that effectively combines on-chip compatibility with high radiative emission rates – a quantum plasmonic launcher. The proposed launchers contain single nitrogen-vacancy (NV) centers in nanodiamonds as quantum emitters that offer record-high average fluorescence lifetime shortening factors of about 7000 times. Nanodiamonds with single NV are sandwiched between two silver films that couple more than half of the emission into in-plane propagating surface plasmon polaritons. This simple, compact, and scalable architecture represents a crucial step towards the practical realization of high-speed on-chip quantum networks.

**Introduction**

Photons are unique carriers of quantum information thanks to their high propagation speed and low decoherence rates. However, photonic quantum technologies require significant resources to compensate for the weak interaction of photons with their environment. One consequence of this weak coupling is the non-determinism of quantum logical operations [1]. These limitations require scalable, chip-compatible single-photon sources emitting at very high rates, ideally in the THz range. On-demand room-temperature single-photon production can be achieved with solid-state single-photon emitters [2-4]. Intrinsically, their emission is non-directional with its rate limited to about 1 GHz [5]. Using various optical nanostructures featuring enhanced light-matter coupling [6-10], one can strongly increase single-photon emission rates and obtain high on-chip collection efficiency [11]. In the proposed approaches, the single-photon rates achievable with dielectric nanostructures is fundamentally limited to the GHz range [12]. Moreover, high quality factors (typically in the range from $10^4$ to $10^6$) [13] of such nanostructures require low-temperature operation so that the quantum emission spectrally fits into the narrowband modes of the photonic resonators.

Plasmonic antennas have emerged as an attractive platform for boosting quantum emission rates. Their unique advantages are the broad bandwidth and the ability to confine light beyond the diffraction limit, leading to giant radiative enhancement factors [14]. Theoretically, plasmonics can lead to single-photon production rates that are several orders of magnitude larger than those offerd by dielectric nanostructures [12]. Leveraging on these ultrafast emission rates, plasmonic nanostructures could enable the on-demand production of indistinguishable photons, even at non-cryogenic temperatures [15]. In this approach, one limitation is that plasmonic materials typically exhibit relatively high ohmic losses. However, using appropriately designed cavity-antenna systems, the quenching rates due to such losses can be kept below the plasmon emission rates. For instance, with the nanoparticle-on-metal structures [16], the plasmon outcoupling to far-field takes place on a time scale comparable to the photon loss rate [17, 18]. Record-breaking performance was demonstrated with emitters coupled to crystalline silver nano-patch antennas [19-21], leading to detected single-photon rates exceeding 35 million counts per second (Mcps).

Leveraging on this progress, , two major stumbling blocks should be overcome to realize ultrafast integrated single-photon sources. First, the device performance is highly sensitive to the relative positions of the dipoles and resonators. The controlled, deterministic fabrication of single-photon sources enhanced by plasmonic nanoantennas [22-29] must attain better precision and repeatability. Second, plasmonic nanoantennas typically feature poorly directional or out-of-plane emission and thus are not directly compatible with on-chip integration [14]. On the other hand, on-chip plasmonic waveguides, such as metal grooves [30], metal/insulator/metal slabs [31], nanowires [32, 33], metal wedge waveguides [34] and dielectric-loaded surface plasmon polariton waveguides [35-37], feature in-plane coupling, but offer limited radiative rate enhancement and suffer from substantial optical propagation losses [38]. A single-photon source simultaneously featuring directional emission and ultrafast operation has still not been realized.

Here, we introduce "quantum plasmonic launchers" (QPLs) as an attractive implementation of plasmon-enhanced on-chip single-photon sources. The proposed structure dramatically enhances the emitter radiative rate and launches in-plane propagating surface plasmon polaritons (SPPs) [39]. The SPP modes themselves can be efficiently coupled into low-loss on-chip photonic waveguides before any significant propagation losses occur [40-42]. In this work, we demonstrate single nitrogen vacancy (NV) centers [43] in nanodiamonds coupled to compact plasmonic launchers (see Fig 1). We first numerically study the total decay rate enhancement and in-plane SPP coupling efficiency for the QPL structure as a function of its salient geometric parameters. Then, we experimentally realize a QPL, recording NV fluorescence lifetimes on the order of 10 ps with emission rate into SPPs accounting for over half of the total radiative rate.

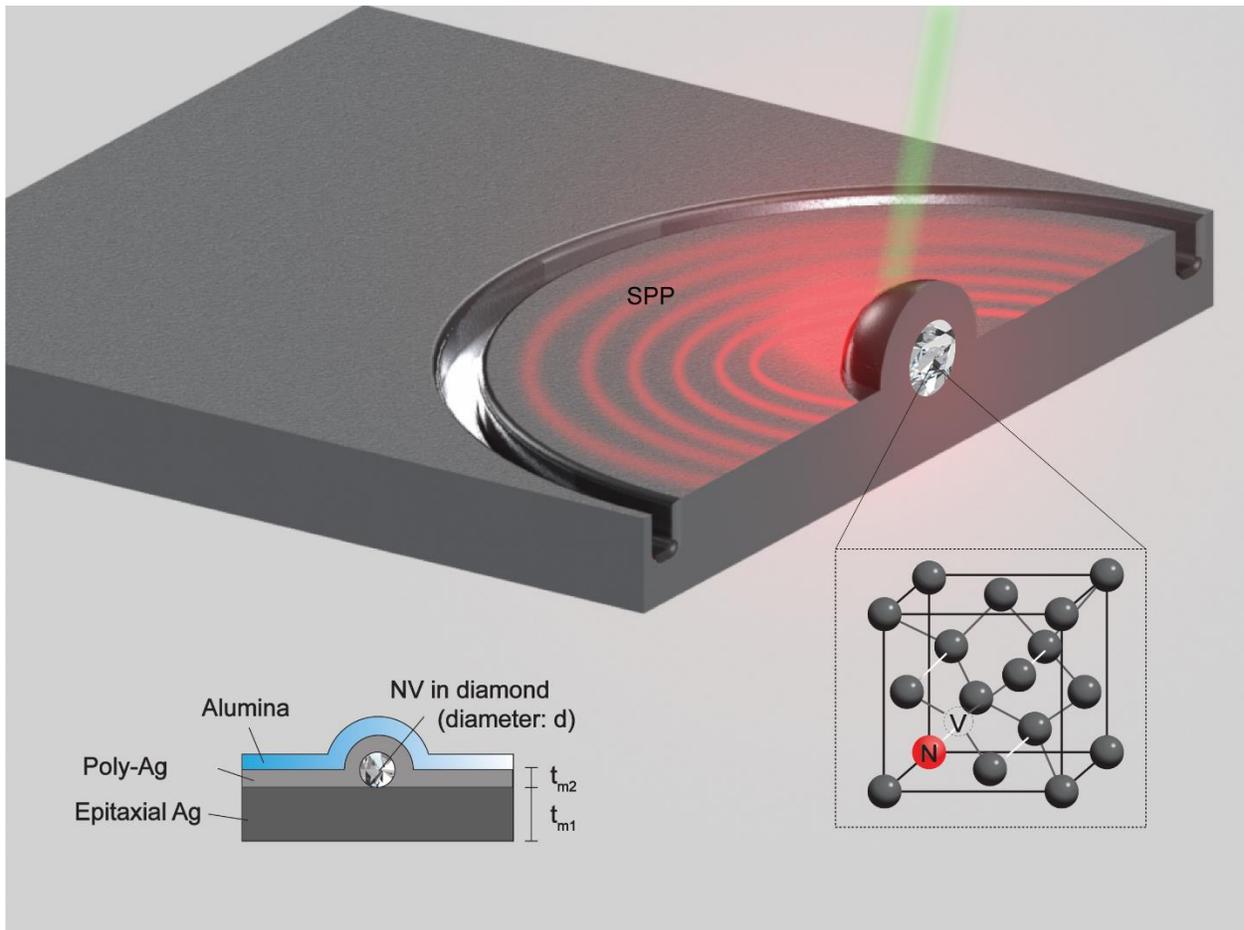

**Fig 1.** Artistic representation of a quantum plasmonic launcher (QPL): a nanodiamond of diameter *d* with a single NV center is placed between an optically thick ($t_{m1}$) and an optically thin ($t_{m2}$) silver films. The structure is coated with a 3 nm thick alumina layer. The NV emission is strongly enhanced and couples preferentially to in-plane surface plasmon modes, making this design compatible with on-chip integration.

**Numerical simulations**

The proposed QPL structure consists of nanodiamonds with single NV centers sandwiched between two silver films of unequal thickness. The resulting structure should promote preferential emission into in-plane surface plasmon modes (see Fig 1). The QPL features a mode volume which is several orders of magnitude smaller than the wavelength cubed, and is limited largely by the volume of the diamond nanoparticle itself. It is therefore expected to exhibit a dramatically enhanced NV total decay rate ($\gamma_{QPL}$) compared to that of a reference NV in a nanodiamond in the dielectric environment, in our case on a glass substrate ($\gamma_0$). The total decay rate enhancement (DRE) is defined as $\text{DRE} = \gamma_{QPL}/\gamma_0$. In turn, $\gamma_{QPL} = \gamma_{FF} + \gamma_{SPP} + \gamma_{NFloss}$, where $\gamma_{FF}$ is free-space photon emission rate, $\gamma_{SPP}$ is surface plasmon launch rate and $\gamma_{NFloss}$ is the local loss rate due to absorption in the immediate vicinity of the emitter. We quantify the performance of the QPL using two parameters. The first parameter is the plasmon-photon emission branching ratio, $\xi = \dfrac{\gamma_{SPP}}{\gamma_{FF} + \gamma_{SPP}}$, representing the fraction of SPPs in the emission from the QPL. The rest of the emission consists of photons radiated into the free space. While being detrimental in the context of the QPL, this free-space leakage allows for the characterization of the emitter in an optical microscope setting. The second parameter is the total plasmon generation efficiency $\beta_{SPP} = \dfrac{\gamma_{SPP}}{\gamma_{QPL}}$, representing the number of plasmons generated per excitation event. It is this parameter that quantifies the "on-demandness" of the QPL by taking into account the rate of local loss $\gamma_{NFloss}$.

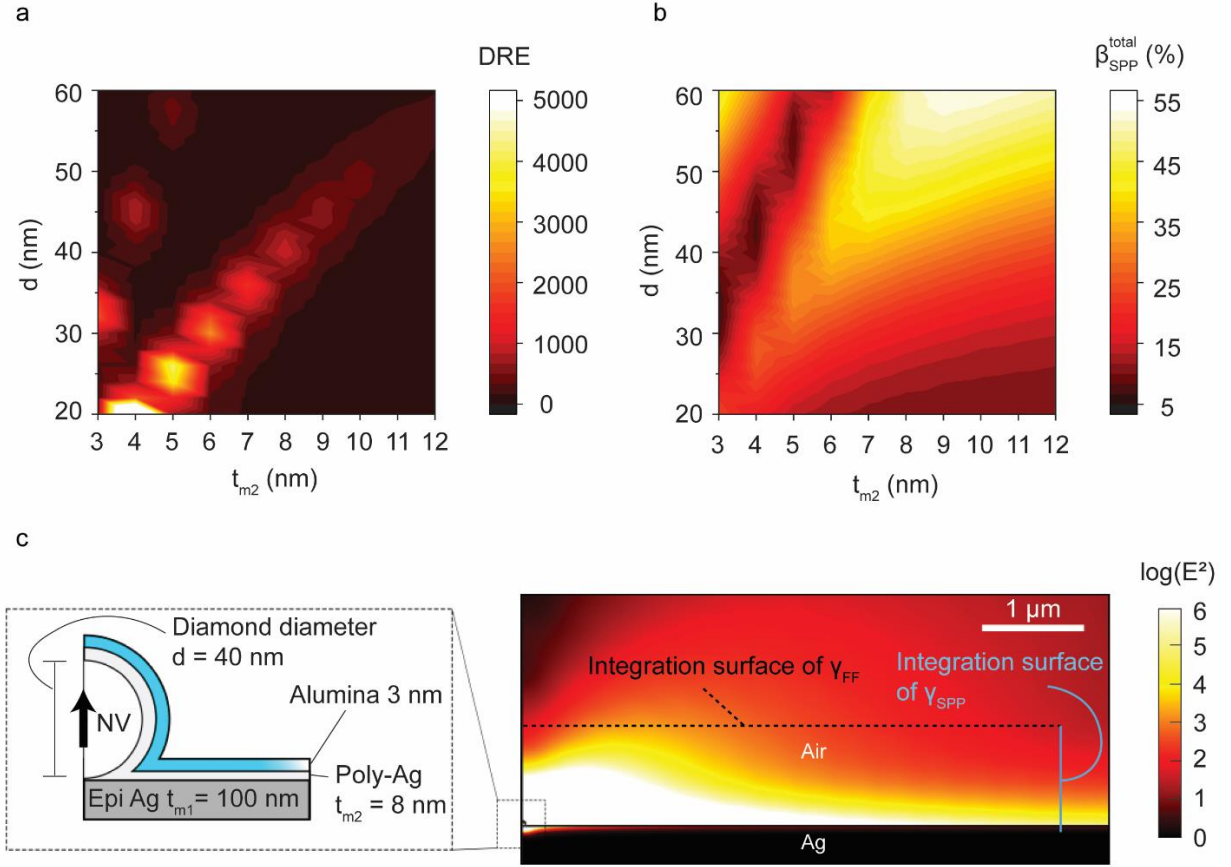

**Fig 2.** Simulated dependence of (a) the total decay rate enhancement (DRE) and (b) the total plasmon generation efficiency ($\beta_{SPP}$) on optically thin ($t_{m2}$) silver films and diamond diameters (*d*). (c) Cross-section of the quantum plasmonic launcher (QPL) and the simulated power flow distribution generated by a single NV center coupled to the QPL.

In the simulated structure, we assume a spherical nanodiamond shape of diameter *d* with an NV center represented by a single vertical dipole in the nanodiamond center. The nanodiamond is sandwiched between the bottom silver film of thickness $t_{m1} = 100 nm$ and an optically thin silver layer on top of thickness $t_{m2} = 8 nm$, overcoated with 3-nm-think layer of alumina. In the experiment described below, we used epitaxial silver to implement the bottom metal film. The top polycrystalline silver film was deposited over the nanodiamond particles. We model the optical characteristics of the two metal films accordingly. The DRE and $\beta_{SPP}$ for a vertically oriented quantum emitter were numerically calculated (for more details, see Supplementary I) by sweeping $t_{m2}$ from 3 nm to 12 nm and $d$ from 20 nm to 60 nm, respectively. The dependences of DRE and $\beta_{SPP}$ on geometric parameters are illustrated in Fig. 2a and 2b. Two distinct families of resonances occur in the parameter space, corresponding to high DRE (Fig. 2a). However, only one of them corresponds to an efficient emission into SPPs. (Fig. 2b). As $d$ increases, so does the cavity volume and the DRE expectedly drops. Considering the proper balance between coupling into plasmons and rate enhancement, we have choosen the design parameters of $d$ = 40 nm and $t_{m2}$ = 8 nm, at which the $\beta_{SPP}$ was calculated to be 32%, while a DRE was maintained at relatively high value of >1000. Figure 2c summarizes the structure's optimal geometrical parameters and shows the plot of the normalized electrical field of the dipole emission for this choice of parameters. The fluorescence power coupled into the SPP mode is 2.8 times larger than that emitted into the far-field (for more details, see supplementary I). These simulation results indicate that even assuming relatively lossy plasmonic materials such as polycrystalline silver, our QPL represents an attractive architecture for realizing on-chip single-photon sources.

**Sample design and characterization**

To validate the QPL concept experimentally, we fabricated two samples: a reference sample A consisting of a bare glass coverslip substrate with dispersed nanodiamonds and a sample B with the NVs in the nanodiamonds coupled to QPLs. We measured the photophysical characteristics of 10 single NV centers in sample A (Supplementary Information, Section III) and 7 single NV centers in sample B (Supplementary Information, Section IV).

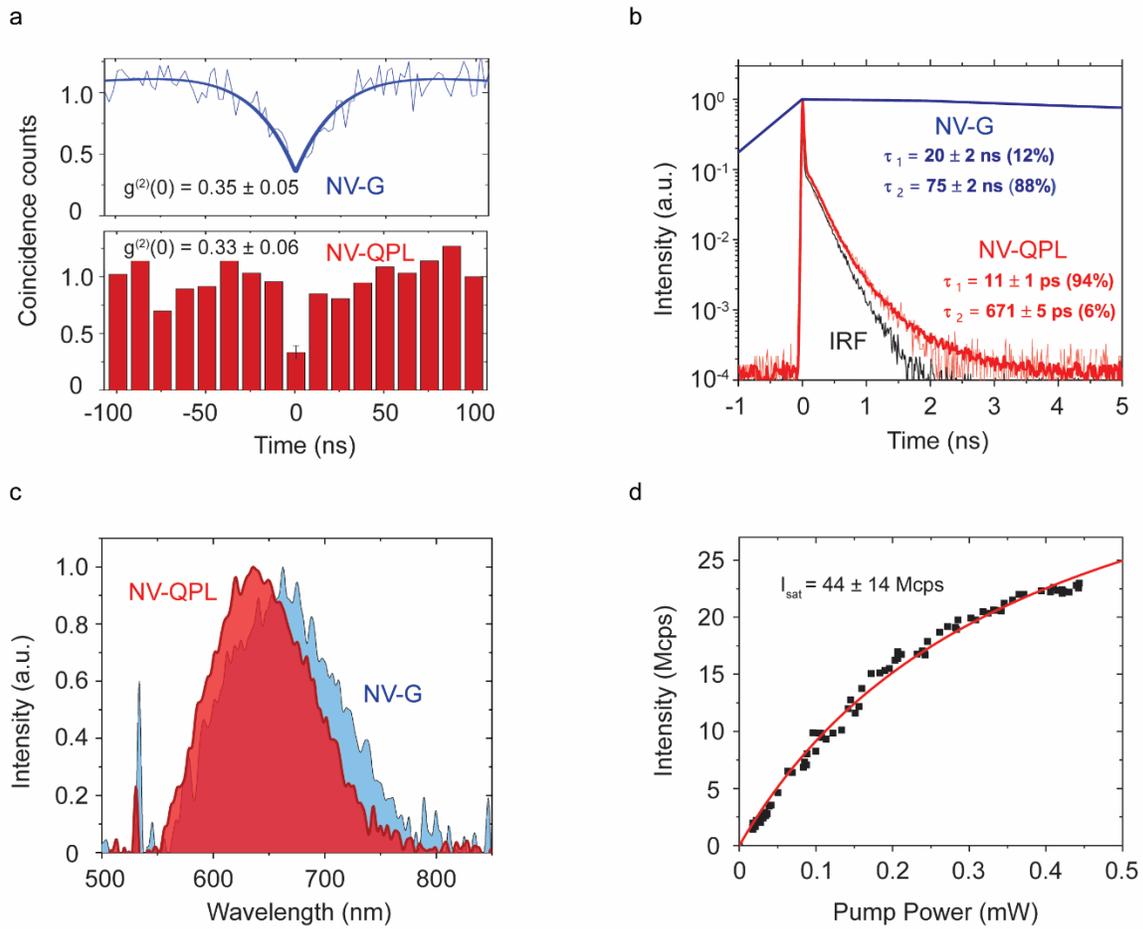

**Fig 3.** Photophysical characterization of the NV-G (blue) and the NV-QPL emitters (red). (a) photon autocorrelation, (b) fluorescence decay curves with the IRF plotted in black for reference, (c) measured photoluminescence spectra, and (d) fluorescence saturation curve.

We selected one emitter exhibiting median fluorescence lifetime values from each of the sample A and the sample B and compared their properties. Henceforth, these emitters are referred to as NV-G and NV-QPL, respectively. Antibunching behavior was characterized by the value of the second-order autocorrelation function $g^{(2)}(t)$ at zero delay, see Fig 3a. While NV-G exhibited a clear antibunching behavior measured using a continuous laser, NV-QPL emission's antibunching feature could not be resolved in time domain using the same setup. Thus, the autocorrelation measurement for NV-QPL was performed using a fs pump laser operating at 1040 nm, doubled to produce an excitation beam at 520 nm. The extracted value of $g^{(2)}(0)$ was 0.33 ± 0.06 for NV-QPL is indicative of a single NV coupled to the QPL. Using pulsed laser excitation, we characterized the total decay rate of the NV-QPL. The time-resolved fluorescence response to the pulse excitation exhibited a faster and a slower decay components, as shown in Fig 3b. The resulting curve was fit with a sum of two exponential functions convoluted with the separately characterized instrument response function (IRF). The decay constants of NV-QPL were $\tau_1 = 11 \pm 1$ ps and $\tau_2 = 671 \pm 5$ ps with intensity weights of 94% and 6%, respectively. The measured fluorescence lifetimes of NV-G were 20 ± 2 ns (12%) and 75 ± 2 ns (88%), leading to a fluorescence lifetime shortening of 6800 ± 800 times based on the ratio of the dominant decay components. Figure 3c compares the spectra of NV-QPL and NV-G emission. Both photoluminescence spectra extend from 570 to 780 nm and significantly overlap, confirming that the enhanced emission resulted from the NV. Fig. 3d shows the background corrected fluorescence rate of the NV-QPL as a function of the CW excitation laser power. The measured fluorescence intensity includes the NV center fluorescence (saturating term), and the background emission (linear term). The linear background fraction was calculated as $r_{bg} = 1 - \sqrt{1 - g^{(2)}(0)}$ [44]. The background contribution was subtracted from the measured fluorescence data. By fitting the background-corrected data, we have obtained a saturated fluorescence detection rate of 44 ± 14 Mcps.

**SPP coupling efficiency**

The QPL structure provides a dramatic enhancement of the total emitter decay rate and a significant increase in the source brightness, sharing these characteristics with recent results based on the use of regular plasmonic antennas. Here, we show that the QPL indeed routes a significant fraction of the emission into in-plane SPPs. To measure the SPP branching ratio $\xi$ and the efficiency $\beta_{SPP}$ of the QPL, we milled a circular trench around a QPL emitter. The trench served as an outcoupler of the SPP-coupled portion of the fluorescence into the far-field, to be collected with the microscope objective (Fig. 4a). The QPL emitter (shown as #2 in supplementary IV) with a nanodiamond height $d = 40 \pm 4$ nm was selected in accordance with the optimal nanodiamond size derived from simulation. After the trench fabrication, we excited the emitter using a CW pump laser operating at 532 nm and recorded fluorescence images of the structure with a CCD camera. The fluorescence image exhibited a bright spot in the center of the QPL structure and a dimmer ring resulting from the portion of the emission coupled into the SPPs, scattered by the trench into the far-field. To compare the intensities collected from the QPL location ($I_{QPL}$) and the circular trench ($I_{ring}$), we have recorded two fluorescence images, shown in Fig. 4b, with exposure times, suitably chosen to obtain substantial, but not saturated pixel intensities at the trench and QPL location respectively. To experimentally obtain the SPP branching ratio $\xi$ and total efficiency $\beta_{SPP}$, we have normalized the measured intensities by the coupling and collection efficiencies as well as taken into account the local loss using the following relations:

$$\xi = \frac{\dfrac{I_{ring}}{\eta_{col}^{ring}\eta_{SPP}^{FF}e^{-r/L}}}{\dfrac{I_{dipole}}{\eta_{col}^{dipole}} + \dfrac{I_{ring}}{\eta_{col}^{ring}\eta_{SPP}^{FF}e^{-r/L}}} \quad (1)$$

$$\beta_{SPP} = \xi\left(1-\beta_{NFloss}\right) \quad (2)$$

Here $\beta_{NFloss} = \gamma_{NFloss}/\gamma_{QPL}$, $\eta_{col}^{dipole} = 57.9\%$ and $\eta_{col}^{ring} = 84.2\%$ stand for the fraction of the local loss, collection efficiency from the QPL and the trench corresponding to the collection solid angle of the air objective (64.2°, 0.9 NA), respectively. The quantity $\eta_{SPP}^{FF} = 30.6\%$ represents the fraction of the SPP power scattered to the far-field by the trench (see supplementary I for calculations of these efficiencies). The factor $e^{-r/L}$ accounts for the experimentally measured propagation loss of the SPPs at the silver-air interface over the distance equal to the trench radius *r* (Supplementary Information, Section V). We used the saturation data from Fig. 3d and the previously characterized efficiency of our setup [19] to estimate the local loss rate in the near field region. These factors are schematically illustrated in Fig. 4c to facilitate the interpretation of Equation (1) and (2).

Our experimentally estimated $\beta_{NFloss}$ value of 98% is higher then the 56.4% obtained from the simulation. Similarly, the values of $\xi$ of 52.1 ± 0.8% and $\beta_{SPP}$ of 1.7 ± 0.1%, are markedly smaller than the simulation values of 73% and 32%, respectively. This discrepancy might stem from the fact that the silver around the trench was exposed to air and degraded rapidly after the FIB processing whereas the rest of the film was protected by the alumina layer. Another contributing factor could be the dipole position inside the nanodiamond. In the numerical simulation, we have assumed the dipole to be in the center of the spherical nanodiamond particle.

In the experiment however, the dipole could be closer to the metal which would dramatically increase local optical losses (see Supplementary Information, Section VI).

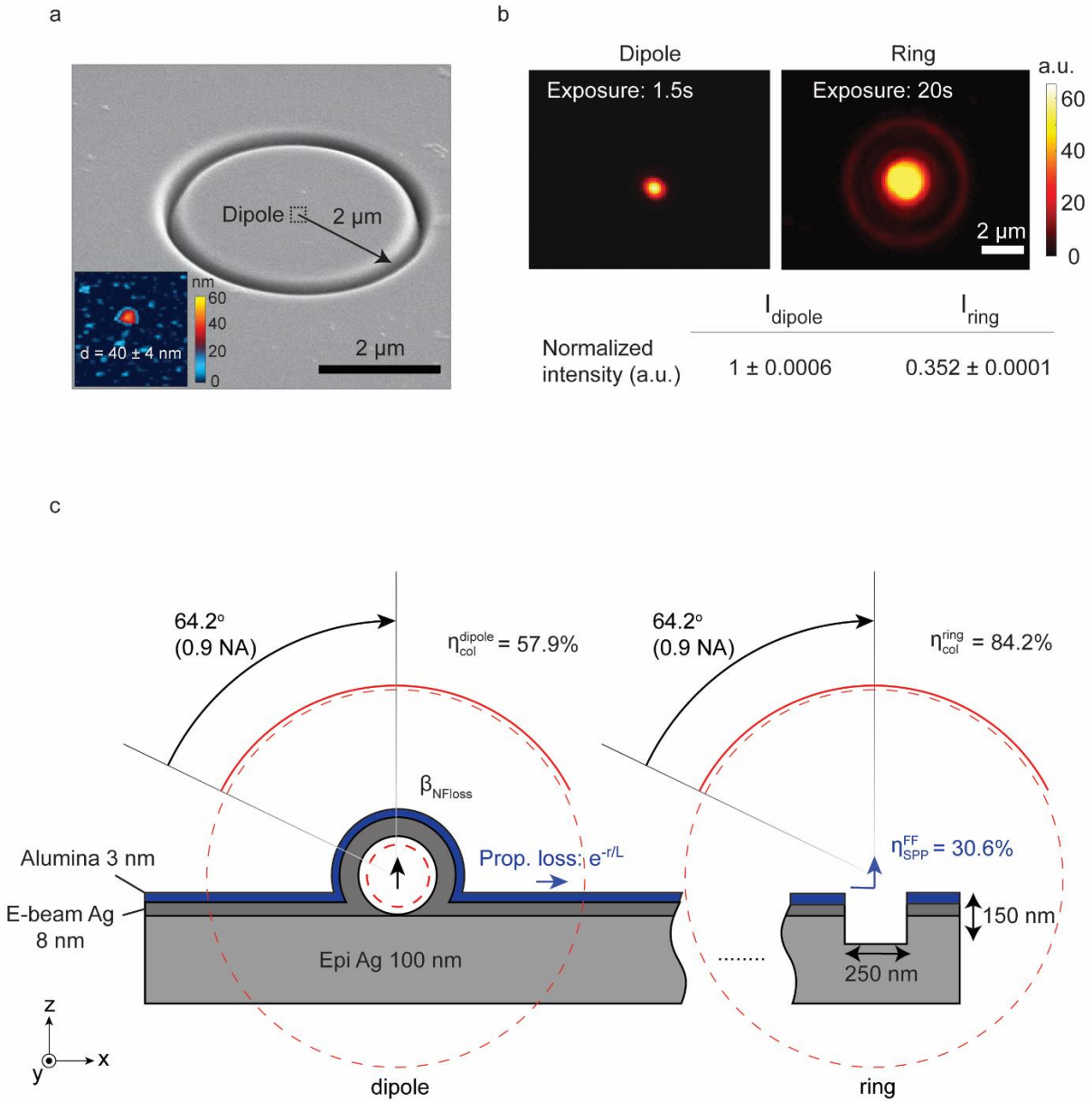

**Fig 4.** (a) Scanning electron micrograph (SEM) image of the circular trench used for SPP outcoupling. Inset in (a) shows the height of the QPL structure measured with AFM. (b) The Optical image recorded with the CCD camera

showing a bright spot in the center from the fluorescence of source and the dimmer ring from scattering of the FIB trench. (c) Schematic illustration of simulated parameters of the QPL and the trench.

**Discussion**

The proposed architecture for quantum plasmonic launchers offers fluorescence lifetime shortening factors of several thousand, far beyond those realized in dielectric loaded waveguides [35, 36], the V-groove system [45] and the chemically synthesized metal nanowires [46]. In addition, the fabrication of the proposed plasmonic launcher requires no lithography. Furthermore, the fabrication process is more scalable and is fully compatible with the on-chip integration of high-speed single-photon sources. At the same time, the plasmonic launcher reaches an SPP branching ratio from quantum emitters, that is similar to that observed in plasmonic waveguide configurations.

In this work, we also measured the total plasmon generation efficiency $\beta_{SPP}$. We found that it is significantly lower than the theoretically simulated value. However, even with the present performance, the QPL is able to generate multi-photon states on-chip at practically siginificant rates (e.g. up to $10^5$ photon triplets per second). Several approaches exist to substantially improve $\beta_{SPP}$. In this initial experiment, the top film was made of polycrystalline silver with relatively high optical losses and roughness compared to the epitaxial silver making up the bottom film. This is a major culprit compromising the total SPP efficiency of the QPL. However, the recently published methods for surface functionalization may reduce the surface roughness by chemically encapsulating the NDs with a thin smooth noble metal shell [47-49]. Furthermore, the morphology and crystallinity of noble metal films can be modified by either thermal [50] or laser-induced annealing [51]. These methods can lead to a more efficient QPL performance.

In the QPL design, the nanodiamond material separates the optical dipole from the metallic surfaces. A certain minimum separation is required to avoid strong non-radiative quenching. However, in our experiments, the location of NV centers within nanodiamonds could not be controlled, leading to excess quenching loss. However this problem can be overcome by using nanodiamonds with NV centers located at the center. The recently demonstrated nanodiamonds grown around single organic precursor molecules [52] provide a promising solution.

The proposed QPLs offer both the high emission rate and the compatibility with on-chip integration. They may be used to efficiently launch single photons into low-loss dielectric waveguides at rates approaching the THz range. For this goal, the SPP mode must be adiabatically converted into a photonic waveguide mode on a length scale shorter than the plasmonic propagation length. This approach makes it possible to interface highly sub-wavelength modes of plasmonic antennas with the modes of an on-chip dielectric waveguide [53, 54]. The QPL and the SPP mode can then be viewed as an impedance matching circuit between a localized plasmon mode and a propagating photonic mode in a dielectric waveguide. Embedding narrowband quantum emitters such as germanium- [36], silicon- [55-57] or tin- [58, 59] vacancy centers in diamonds into this launcher opens the possibility of realizing high-speed integrated quantum optical networks operating at or close to room temperature [15].

**Methods**

**Fabrication.** Each nanodiamond (diameter: 40 ± 20 nm) contained one to four NVs (Adamas Nano). Reference sample A was fabricated by drop-casting the nanodiamonds on a glass coverslip substrate with refractive index $n = 1.525$. Sample B was fabricated by depositing a 100 nm epitaxial silver on a MgO substrate. Nanodiamonds were drop-casted on the silver layer, and overcoated by e-beam evaporated silver and alumina with 8 nm and 3 nm thicknesses, respectively, in the same deposition chamber without breaking the vacuum. To measure the SPP coupling efficiency of this structure, we fabricated a circular trench around the emitter using focused ion beam milling. The radius, width, and depth of the circular trench were 2000 nm, 250 nm, and 150 nm, respectively.

**Characterization.** Experiments were performed on a home-built scanning confocal microscope with a 50 μm pinhole based on a commercial inverted microscope body (Nikon Ti–U). The optical pumping in the continuous wave (CW) experiments was by a 200 mW continuous wave 532 nm laser (Shanghai Laser Century). Femtosecond pulsed autocorrelation measurements were performed using a compressed tunable mode-locked laser with an 80 MHz repetition rate (Mai Tai DeepSee, Spectra Physics). The laser was set to operate at a wavelength of 1040 nm, and its output was frequency doubled to obtain a 520 nm emission. Lifetime measurements were performed with this SPC-150 system while exciting an NV center with the doubled Mai Tai DeepSee with a nominal 100 fs pulse width and a 514 nm fiber-coupled diode laser with a nominal 100 ps pulse width and adjustable repetition rate in the 2–80 MHz range (BDL-514-SMNi, Becker & Hickl). The excitation beam was reflected off a 550 nm long-pass dichroic mirror (DMLP550L, Thorlabs), and a 550 nm long-pass filter (FEL0550, Thorlabs) was used to filter out the remaining pump power. After passing through the pinhole, two avalanche detectors with a 30 ps time resolution and 35% quantum efficiency at 650 nm (PDM, Micro-Photon Devices) were used for single-photon detection during scanning, lifetime, and autocorrelation measurements. An avalanche detector with 69% quantum efficiency at 650 nm (SPCM-AQRH, Excelitas) was used for saturation measurements. Time-correlated photon counting was performed by an acquisition card with a 4 ps internal jitter (SPC-150, Becker & Hickl).

To perform the wide-field fluorescence measurements, we implemented a CCD camera (Atik 414EX, Atik Cameras) accessible through additional collection channel employing a flip mirror. An additional 550 nm long-pass filter was put in front of the CCD camera.

**Author Contributions**

S.B. and C.-C.C designed the experiments. C.-C.C. fabricated the devices, performed the quantum emitter characterization and wrote the initial draft of the paper. O.A.M., C.-C.C. and D.W. carried out the numerical simulations. X.X. and C.-C.C. fabricated a circular trench around the emitter using focused ion beam milling. S.B. and A.S.L. built the experimental setup. S.S. and D.S. grew the epitaxial silver film and carried out the ellipsometry characterization of silver substrates. C.-C.C., S.B., O.A.M., D.S., D.W., A.S.L. and A.B. contributed to writing the manuscript and discussed the results at all stages. A.B. and V.M.S. supervised the project. All authors have given approval to the final version of the manuscript.

**Acknowledgements**

This work was partially supported by the U.S. Department of Energy, Office of Basic Energy Sciences, Division of Materials Sciences and Engineering under Award DE-SC0017717 (S.B.)

and the Office of Naval Research (ONR) DURIP Grants No. N00014-16-1-2767 and N00014-17-1-2415(equipment grants used to purchase the scanning confocal microscope, lasers, detectors and single-photon counting capability used in this work). A.V.K. acknowledges the DARPA/DSO EXTREME, Award HR00111720032 (numerical modeling and simulations).

**Competing interests**

The authors declare no competing financial interests

**Data availability**

The data that support the plots within this paper are available from the corresponding author upon request.

# Supplementary Information

# Chip-compatible quantum plasmonic launcher

Chin-Cheng Chiang, Simeon I. Bogdanov, Oksana A. Makarova, Xiaohui Xu, Soham Saha, Deesha Shah, Di Wang, Alexei S. Lagutchev, Alexander V. Kildishev, Alexandra Boltasseva, and Vladimir M. Shalaev

School of Electrical & Computer Engineering and Birck Nanotechnology Center, Purdue University, West Lafayette, IN 47907, USA

Purdue Quantum Science and Engineering Institute, Purdue University, West Lafayette, IN 47907, USA

**I. Simulation**

All full-wave 3D electromagnetic numerical simulations were performed using the finite-element frequency domain method with commercial software (Comsol Multiphysics, Wave Optics Module). In the simulation, the optical emitter was modeled as an AC current density inside a 2 nm diameter sphere enclosed by a diamond shell of diameter 40 nm with refractive index $n = 2.42$. The emitter's wavelength was fixed at 685 nm. The emitter had a vertical dipole orientation to provide symmetry so that the 3D model was treated as an axisymmetric 2D problem. The emitter was placed between a 100 nm epitaxial Ag layer and an 8 nm e-beam Ag film, followed by capping 3 nm alumina with refractive index $n = 1.74$. The thickness of alumina was fixed to 3 nm to achieve the optimal DRE, $\beta_{SPP}^{emisson}$ and $\beta_{SPP}^{total}$. We carried out the ellipsometry characterization for both epitaxial Ag and e-beam Ag (Supplementary Information, Section II). The total decay rate $\gamma_{QPL}$ is calculated as the surface integral of total power flow **P** through a 3 nm radius spherical surface $\Omega_{in}$ encapsulating the emitting dipole within the nanodiamond volume (Fig S1): $\gamma_{QPL} \propto \int_{\Omega_{in}} \mathbf{P} \cdot \mathbf{dS}$. The SPP decay rate $\gamma_{SPP}$ is calculated as the surface integral of total power flow **P** through the surface $\Omega_{SPP}$: $\gamma_{SPP} \propto \int_{\Omega_{SPP}} \mathbf{P} \cdot \mathbf{dS}$. The loss rate $\gamma_{NFloss}$ is determined as the total work performed by the electric field on the free charges in the metal parts occupying the volume $V_m$: $\gamma_{NFloss} \propto \int_{V_m} \mathbf{J} \cdot \mathbf{E} dV$. The ratio of local loss from the silver cap $\gamma_{loss}^{cap} / \gamma_{QPL} = 56.4\%$. We normalized all the decay rates by the spontaneous emission decay rate of a dipole on the glass ($\gamma_0$).

The collection efficiency $\eta_{col}$ is calculated as the ratio of the far-field electric fields squared integrated over a spherical outer surface $\Omega_{out} - \Omega_{SPP}$, and the portion of that surface $\Omega_{col}$ corresponding to the collection solid angle of the air objective (64.2°, 0.9 NA). The quantity $\gamma_{SPP}^{FF}$ represents the far-field decay rate for SPP out-coupling through the FIB ring. The simulation shows that 30.6% of input power will convert into far-field decay when SPP is hitting the ring, i.e. $\eta_{SPP}^{FF}$ = 30.6%.

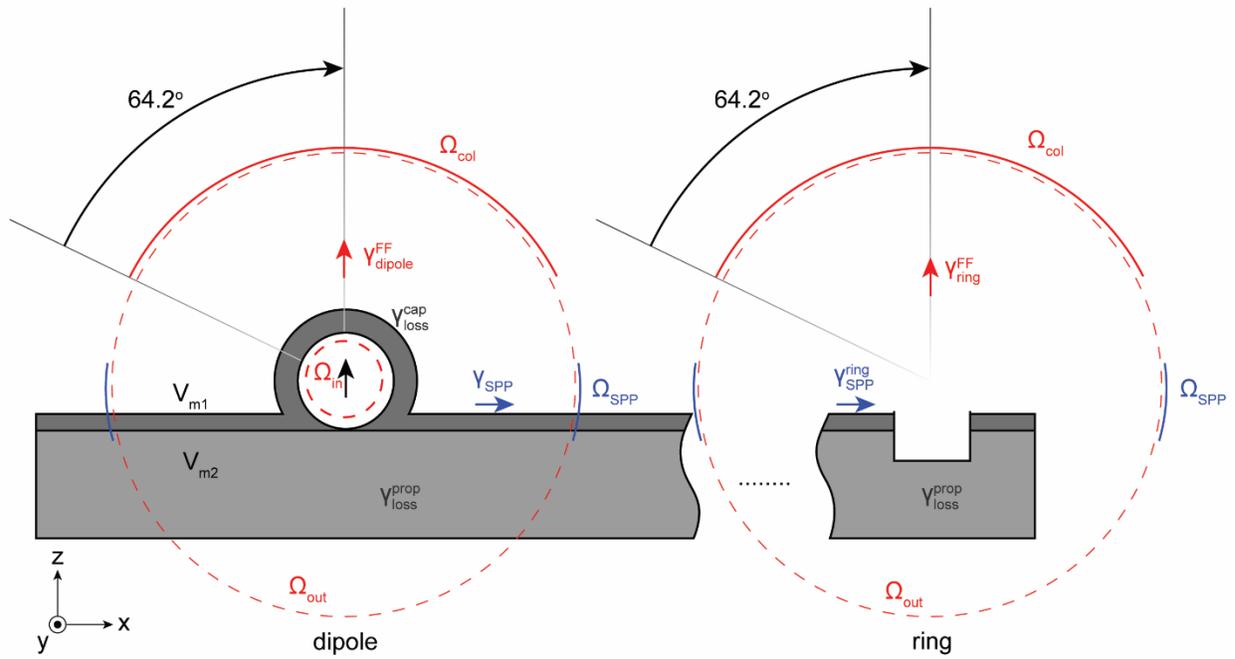

**Fig S1.** Schematic illustration of integration volumes and surfaces used for simulating efficiency parameters of the plasmonic launcher.

The summary of numerically obtained decay rates ($\gamma_{QPL}$, $\gamma_{loss}^{cap}$, $\gamma_{SPP}$, and $\gamma_{FF}$) is presented in Table S1. The calculated collection of efficiencies is summarized in Table S2.

| Plasmonic launcher | Normalized decay rate | (%) |
|---|---|---|
| $\gamma_{QPL}/\gamma_0$ | 977.7 | 100 |
| $\gamma_{loss}^{cap}/\gamma_0$ | 551.6 | 56.4 |
| $\gamma_{SPP}/\gamma_0$ | 313.7 | 32.0 |
| $\gamma_{FF}/\gamma_0$ | 113.5 | 11.6 |

**Table S1.** Decay rates summary table: the decay rates of dipole emission, local loss to EB metal cap, SPP, and far-field.

| | Efficiency (%) |
|---|---|
| $\eta_{SPP}^{FF}$ | 30.6 |
| $\eta_{col}^{dipole}$ | 57.9 |
| $\eta_{col}^{ring}$ | 84.2 |

**Table S2.** Efficiency summary table: the SPP outcoupling efficiency and the collection efficiencies of the dipole and ring.

## II. Silver optical parameters

The polycrystalline silver was deposited using an e-beam evaporator (Leybold) at a pressure of $2\cdot 10^{-6}$ Torr. First, an adhesion layer of Ti (5 nm) was deposited on <100> Si substrate followed by a 50 nm layer of Ag. The sample was characterized by the variable angle spectroscopic ellipsometer (V-VASE, J.A. Woollam) to retrieve the optical properties of the silver. The wavelength used in simulations is 685 nm, and the dielectric permittivity of the substrate at this wavelength is $\varepsilon = -21 + i1.2$.

The epitaxial silver substrate was deposited using reactive magnetron sputtering at a base pressure of $5\cdot 10^{-8}$ Torr. First, an adhesion layer of TiN (4 nm) was deposited on a MgO substrate followed by a 100 nm layer of Ag. The sample was characterized by V-VASE to retrieve the optical properties of the epitaxial silver substrate. The dielectric permittivity of the epitaxial silver substrate at 685 nm wavelength is $\varepsilon = -21 + i0.4$.

## III. Statistics of reference emitters: single NVs on the glass substrate

In the control experiment, we characterized and measured the photophysical properties of 10 nanodiamonds containing single NVs that were dispersed over the glass substrate using an oil objective with NA = 1.49. Only emitters with antibunching characterized by $g^{(2)}(0) < 0.5$ were included in the statistics. We only took the dominant part of the lifetime values and summarized the distribution in Fig S2.

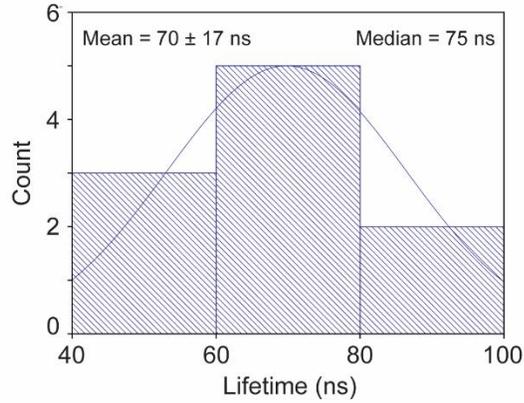

**Fig S2.** Statistical distributions of the fluorescence lifetime of NV centers in nanodiamonds dispersed over the glass substrate and characterized with an oil objective with NA = 1.49, yielding a lifetime distribution of 70 ± 17 ns.

## IV. Additional emitters coupled to the QPL on the epitaxial silver substrate

#1

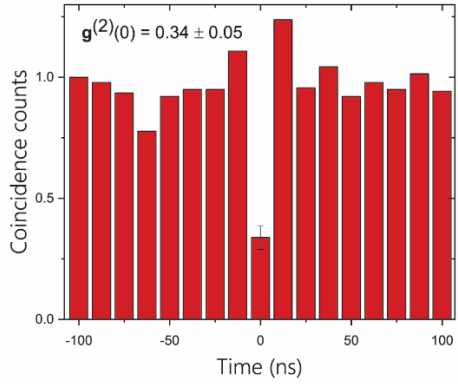
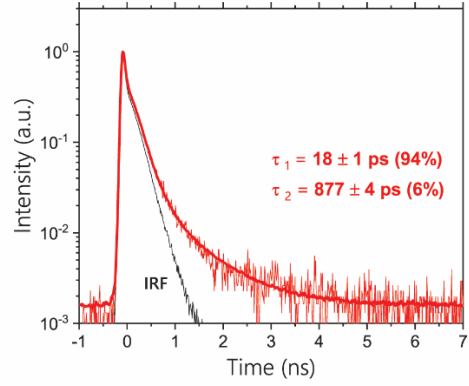

#2

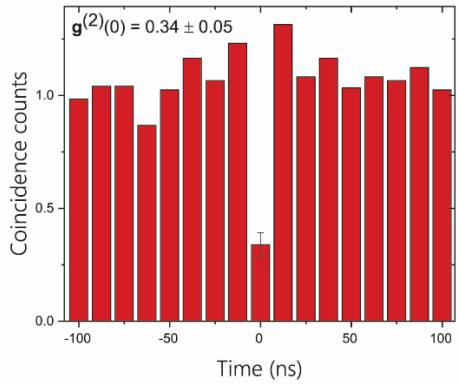
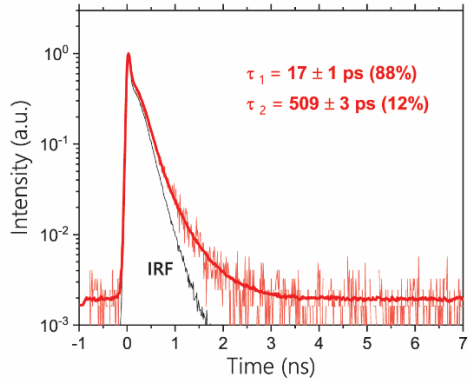

#3

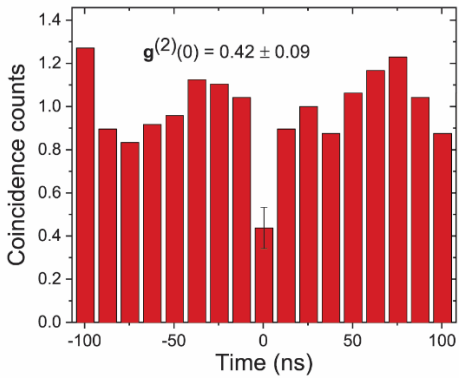
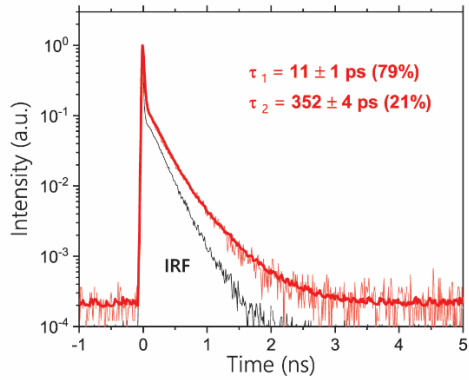

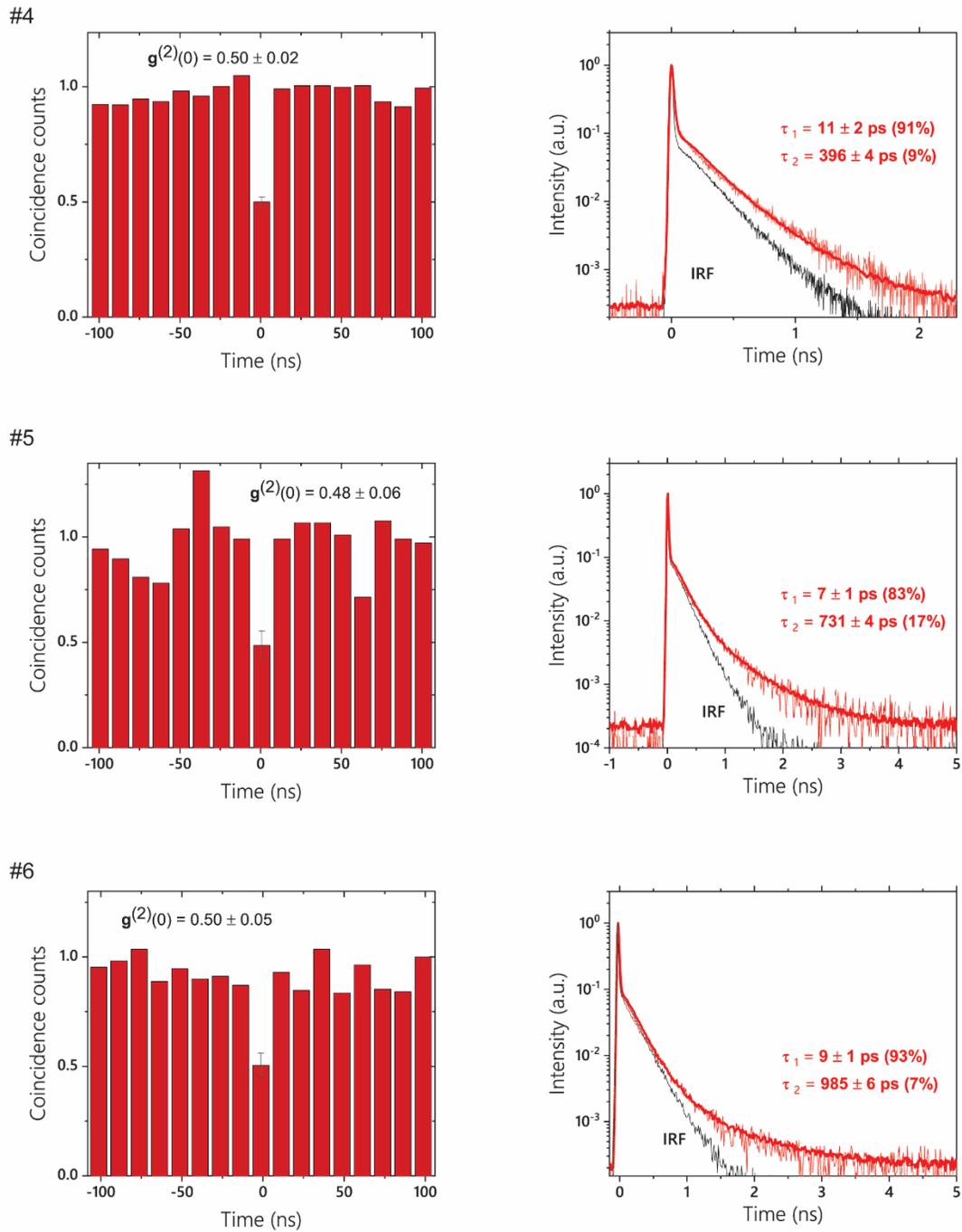

**Fig S3.** Autocorrelation measurements (left column) and fluorescence decays (right column) for several NV center emitters enhanced by the QPL. The QPL samples were characterized using an air objective with NA = 0.9.

## V. Propagation length

To measure the SPP propagation length, we utilized the focused ion beam (FIB) to make several circular trenches around an NV. We compared the attenuation of fluorescence signals at rings with different radii and their corresponding propagation distances from the NV. The propagation length was estimated as an exponentially decaying function of distance to confocal spot, $I = I_0 e^{-x/L}$. Here $x$ is the distance between the trench and confocal spot, $L$ is the propagation length of SPP mode supported by the plasmonic launcher, $I$ is the measured intensity of the emission, and $I_0$ is the original intensity of the emission coupled with SPP mode by an NV. We assumed symmetric coupling in all directions, uniform losses across the silver substrate, and the same out-coupling efficiency at all trenches. We recorded the intensities and corresponding distances to the confocal spot and plotted the results in Fig S5. Using the exponential fit, we obtained a propagation length ($L$) of 6.35 μm ± 0.48 μm.

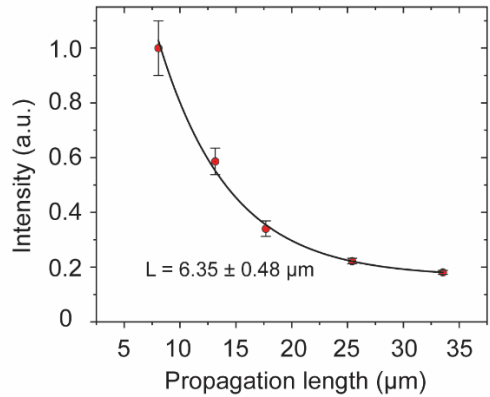

**Fig S4.** Experimental data of the propagation length measurements (red dots) with the curve of the exponential fit (black solid line).

# VI. Simulation results for various dipole positions

Distance dependence of the DRE and $\beta_{SPP}^{total}$ for QPL structure with a vertically oriented dipole is studied. The dipole is moved upward from the surface of the epitaxial silver along the $z$-axis (Fig S5a). The optimal position of the dipole is at the center of the nanodiamond because proximity of the dipole to the metal surface leads to fluorescence quenching and makes the dipole dim (Fig S5b).

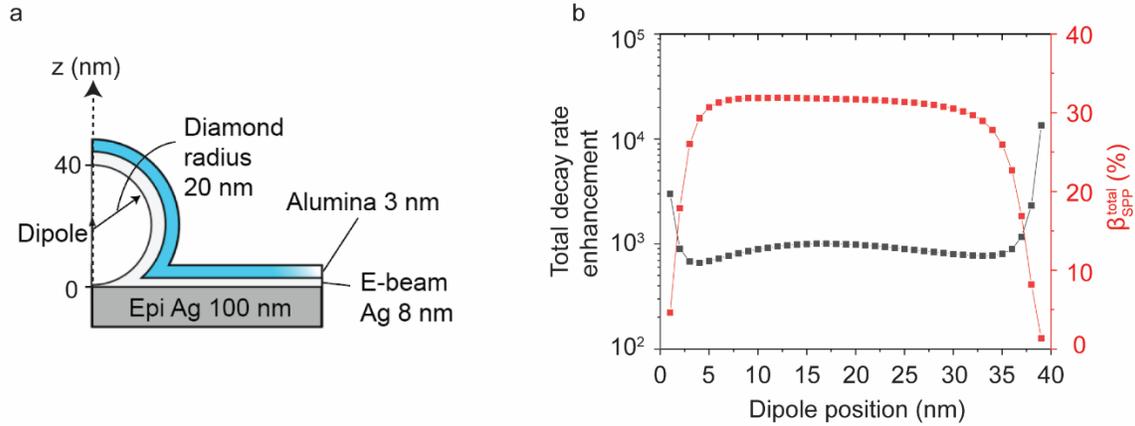

**Fig S5.** (a) The dipole position is measured along the $z$-axis from the surface of the epitaxial silver. (b) Dependence of DRE and $\beta_{SPP}^{total}$ vs. the distance between the dipole and the metal films.

## VII. Surface roughness of the epitaxial silver and E-beam silver

The quality of epitaxial Ag (thickness of 100 nm) was confirmed by atomic force microscopy (AFM). The epitaxial Ag has a root mean square (RMS) roughness of 337 pm. We further deposited 5 nm silver by e-beam evaporation to check the uniformity of the top thin film. The evaporated film had an RMS roughness of 565 pm, which was comparable to the epitaxy Ag (Fig S6a), indicating that the evaporated Ag film was continuous on top of the epitaxy Ag. The NV-QPL structures were assembled by evaporating silver and alumina on the NVs shown in Fig. S6b.

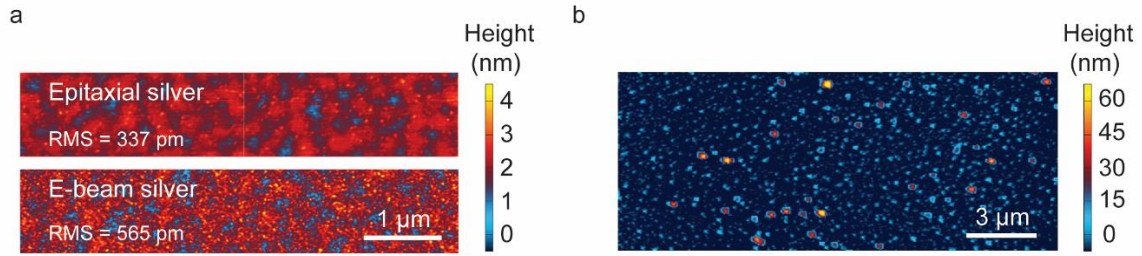

**Fig S6.** (a) The roughness of epitaxial silver and E-beam silver measured with AFM. (b) The AFM image of NV-QPL structures.

## VIII. Fast decay component of NV-QPL

In order to confirm that the fast decay component indeed results from the NV-QPL, we measure the autocorrelation of the NV-QPL emission using different pump lasers. The NV-QPL emission's antibunching cannot be resolved by the CW autocorrelation measurement (Fig S7a). However, in the pulsed excitation regime, using a femtosecond laser with a pulse duration ($t_{pulse}$) below 1 ps, the antibunching appears clearly with a $g^{(2)}(0)$ of $0.33 \pm 0.06$ (Fig S7b).

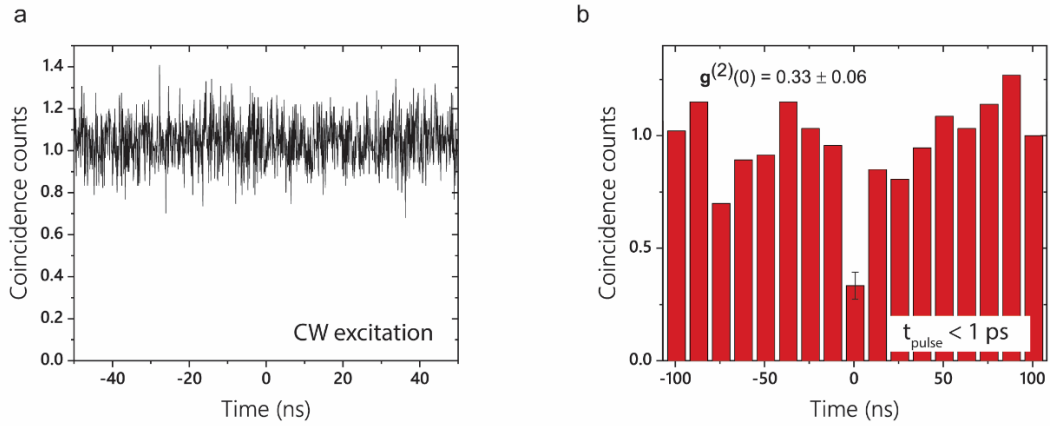

**Fig S7.** Autocorrelation measurements of the NV-QPL. (a) Under continuous wave excitation, the antibunching cannot be measured. (b) Under short pulse excitation with $t_{pulse}$ < 1 ps, the antibunching appears clearly.